# Switchable Altermagnetism in a Layered van der Waals Metal-Organic Framework Driven by Spin-Crossover

Diego López-Alcalá[1], Alberto M. Ruiz[1], Andrei Shumilin[1] and José J. Baldoví[1,]*

[1]Instituto de Ciencia Molecular, Universitat de València, Catedrático José Beltrán 2, 46980 Paterna, Spain

Dynamical control of altermagnetism is a key requirement for translating its unique spin-dependent functionalities into practical spintronic devices, yet effective switching mechanisms remain largely unexplored. Here, we open an unprecedented, versatile and programmable route based on spin-crossover to switch altermagnetism on demand in molecular materials. Using density functional calculations, we demonstrate an altermagnetic ground state in the layered van der Waals MOFs $MnX_2(tdz)_2$ (X = Cl, Br; tdz = thiadiazole), stabilized by anisotropic interlayer exchange interactions, which gives rise to a characteristic d-wave momentum-space spin splitting and an associated spin-splitter transport response. Our findings reveal that under hydrostatic pressure, a high-spin to low-spin transition reconfigures the Mn d-orbital occupation, thus modifying the magnetic exchange network, and stabilizing a different antiferromagnetic ground state whose symmetry suppresses the nonrelativistic spin splitting. Crucially, spin-crossover switches altermagnetism not by directly altering the electronic structure, but by changing the symmetry of the magnetic ground state. These results establish molecular spin-crossover altermagnets as a platform for externally reconfigurable spintronic devices.

Altermagnetism has recently emerged as an unconventional magnetic phase in which collinear antiferromagnets exhibit sizable nonrelativistic spin splitting while preserving zero net magnetization.[1,2] Governed by crystal symmetry, this spin splitting enables spin-dependent transport and current-induced spin functionalities without relying on spin–orbit coupling (SOC).[3] While numerous altermagnetic (AM) materials have recently been identified, achieving dynamic control over their magnetic symmetry and spin splitting remains a major challenge toward practical spintronic applications.[4]

Metal-organic frameworks (MOFs) emerge as an attractive platform to address this challenge owing to their exceptional chemical versatility and modular nature, which enable the rational engineering of crystal symmetry, magnetic exchange interactions and electronic structure.[5–7] Moreover, MOFs exhibit a broad range of electronic and magnetic functionalities, including room-temperature magnetism, topological magnetic phases and chemically tunable magnetic interactions.[8–13] Within this context, coordination networks have recently been proposed as promising AM materials.[14] These span from architectures at the 2D limit to bulk systems.[15–20] However, while these studies considerably expand the family of altermagnets through chemical design, the resulting magnetic properties remain essentially static, precluding dynamic manipulation of the magnetic symmetry and the associated spin splitting.

Among the most distinctive functionalities of magnetic molecular materials is spin-crossover (SCO), which enables reversible changes in the electronic configuration and magnetic state under external stimuli such as pressure, temperature or light.[21,22] Coupling SCO with altermagnetism therefore represents a promising strategy to dynamically manipulate magnetic symmetry and switch nonrelativistic spin splitting on demand.[23,24] Nevertheless, no molecular material has yet been shown to intrinsically couple SCO and altermagnetism, leaving switchable AM functionalities in molecular materials unrealized.

In this work, we demonstrate that SCO provides an effective strategy to switch altermagnetism in molecular materials. We investigate the layered van der Waals MOFs $\mathbf{MnX_2(tdz)_2}$ (X = Cl, Br; tdz = thiadiazole) using density functional theory calculations (DFT). Our simulations reveal an AM ground state stabilized by anisotropic interlayer exchange interactions, which gives rise to a characteristic d-wave spin splitting along the stacking direction. Then, we apply hydrostatic pressure and, interestingly, an SCO transition that reconfigures the Mn *d*-orbital occupation takes place, which changes the magnetic ground state, accompanied by a transition from an AM to a conventional antiferromagnetic (AFM) phase.

The chemical structure of our case study, i.e. **$MnX_2(tdz)_2$**, belongs to the orthorhombic *Cmce* (64) space group, in which the $Mn^{2+}$ centers adopt slightly distorted octahedral environments (Figure 1). These are coordinated by axial halide ligands and equatorial tdz molecules. The alternating orientation of neighboring tdz ligands generates two crystallographically inequivalent Mn sublattices, while preserving the global crystal symmetry. This provides the structural framework required for symmetry-protected altermagnetism. This layered topology closely resembles that of the experimentally synthesized NCS analogue,[25] which supports its structural feasibility, and periodic DFT studies of Mn(II) MOFs have previously addressed their electronic structure and magnetic exchange interactions.[26] Furthermore, ab initio molecular dynamics simulations confirm the dynamical stability of both **$MnCl_2(tdz)_2$** and **$MnBr_2(tdz)_2$**, without structural reconstruction (Figure S6 and S7).

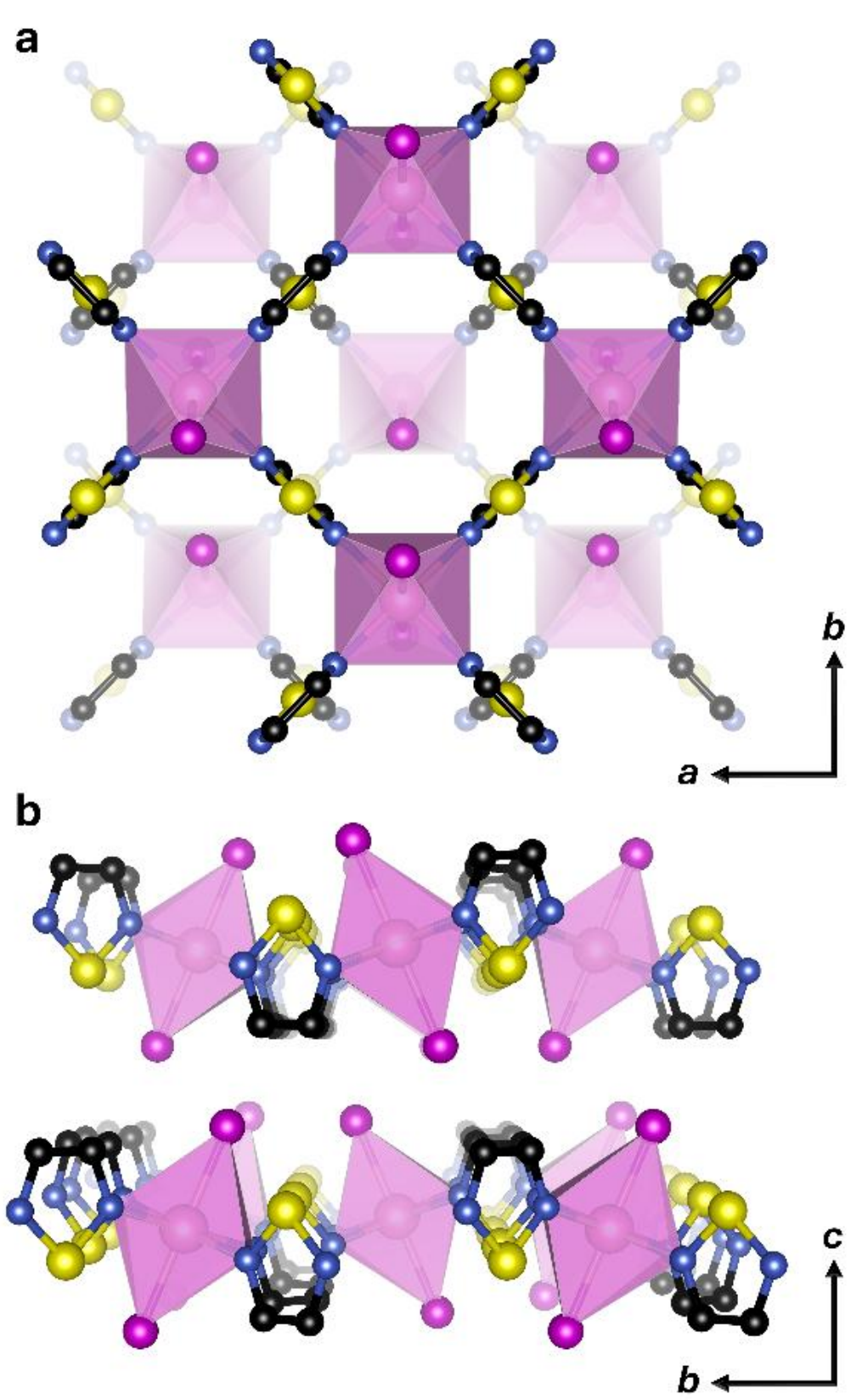

**Figure 1. a** Top and **b** side view of **$MnCl_2(tdz)_2$** structure. Color code: pink (Mn), purple (Cl), yellow (S), blue (N), black (C). H atoms have been omitted for clarity.

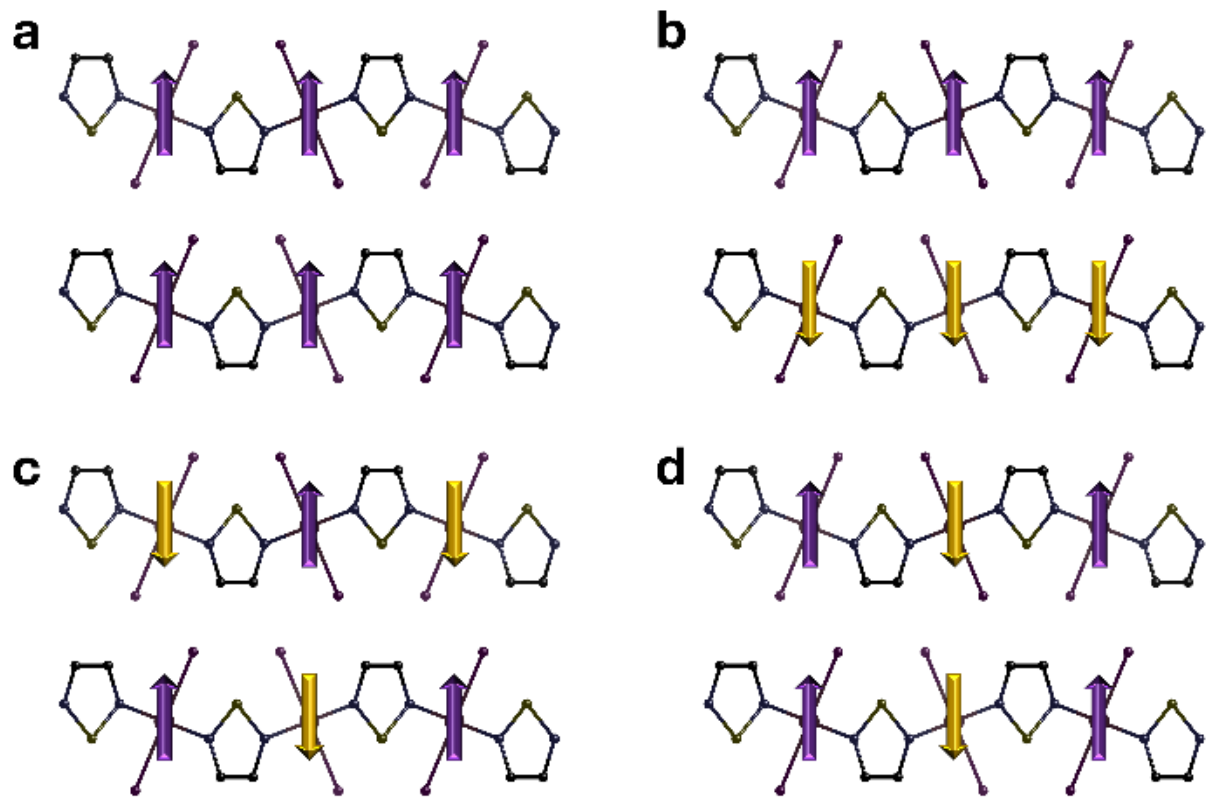

**Figure 2.** Spin orientation of **a** FM, **b** $AFM_1$, **c** $AFM_2$ and **d** $AFM_3$ configurations. Color code: purple(orange) represents spin up(down).

The $Mn^{2+}$ nodes display a calculated magnetic moment of 4.52 $\mu_B$, which is consistent with a high-spin (HS) $d^5$ magnetic configuration ($t_{2g}^3e_g^2$, S = 5/2). Superexchange mediated by the tdz ligands gives rise to several competing magnetic configurations (Figure 2), whose distinct symmetry relationships determine whether altermagnetism is allowed.

The symmetry relation between opposite-spin sublattices is decisive for altermagnetism. While the $AFM_1$ and $AFM_2$ configurations preserve the $[C_2 \parallel i]$ symmetry operation, which connects opposite-spin sublattices and therefore forbids altermagnetism, the $AFM_3$ configuration retains only the $[1 \parallel i]$ symmetry (Supplementary Note 1), fulfilling the symmetry conditions required for an AM ground state.[2]

Our calculations identify $AFM_3$ as the lowest-energy magnetic configuration for both **$MnCl_2(tdz)_2$** and **$MnBr_2(tdz)_2$** (Table 1), demonstrating that the energetically preferred magnetic state is precisely the one compatible with altermagnetism. The stabilization of $AFM_3$ originates from the dominant AFM superexchange expected for HS $Mn^{2+}$, where the partially occupied $t_{2g}$ and $e_g$ orbitals strongly favor AFM coupling over ferromagnetic alignment. Therefore, the existence of altermagnetism in **$MnX_2(tdz)_2$** relies on the stability of the $AFM_3$ spin arrangement. Numerical errors in relative magnetic energies are below 0.006 meV/Mn (Table S1).

**Table 1.** Calculated energy of each magnetic configuration in $\mathbf{MnX_2(tdz)_2}$ in meV/Mn atom.

| | DFT Functional | FM | $AFM_1$ | $AFM_2$ | $AFM_3$ |
|---|---|---|---|---|---|
| $\mathbf{MnCl_2(tdz)_2}$ | PBE+U | 5.61 | 4.47 | 0.77 | 0 |
| | HSE06 | 4.48 | 4.03 | 0.42 | 0 |
| $\mathbf{MnBr_2(tdz)_2}$ | PBE+U | 4.86 | 3.44 | 1.02 | 0 |
| | HSE06 | 3.80 | 3.16 | 0.62 | 0 |

Having established $AFM_3$ as the symmetry-allowed magnetic ground state, we next examine its electronic signature (Figure 3a and S8). $\mathbf{MnX_2(tdz)_2}$ is a semiconductor with a direct band gap of 1.52 eV (1.66 eV in $\mathbf{MnBr_2(tdz)_2}$). As the projected density of states reveals (Figure 3b), the valence-band maximum is dominated by localized Mn *d* states, whereas the conduction-band minimum mainly originates from tdz orbitals, indicating a metal-to-ligand charge-transfer gap.[27,28]

Consistent with the symmetry of the $AFM_3$ ground state, the electronic structure exhibits a highly anisotropic spin splitting in momentum space. While the electronic bands remain strictly spin degenerate along the in-plane crystallographic directions, finite spin splitting emerges exclusively along the stacking direction, allowing the characteristic alternating T–Γ–T′ dispersion expected for an altermagnet (Figure 3c). The corresponding spin-splitting map further reveals the characteristic *d*-wave anisotropy within the *bc* plane, with two nodal directions separating regions of opposite spin polarization (Figure 3d).[1] Our hybrid functional calculations further confirm that the observed spin splitting is an intrinsic electronic property of $\mathbf{MnX_2(tdz)_2}$ rather than a functional-dependent artifact (Figure S10).

The persistence of this spin splitting upon inclusion of SOC confirms its purely nonrelativistic origin (Figure S11). The Br analogue exhibits the same electronic behavior despite the larger interlayer separation and enhanced relativistic effects associated with the heavier Br atoms, demonstrating the robustness of the AM ground state. More importantly, the pronounced momentum-space anisotropy of the spin splitting is expected to directly determine the spin-transport response of $\mathbf{MnX_2(tdz)_2}$.

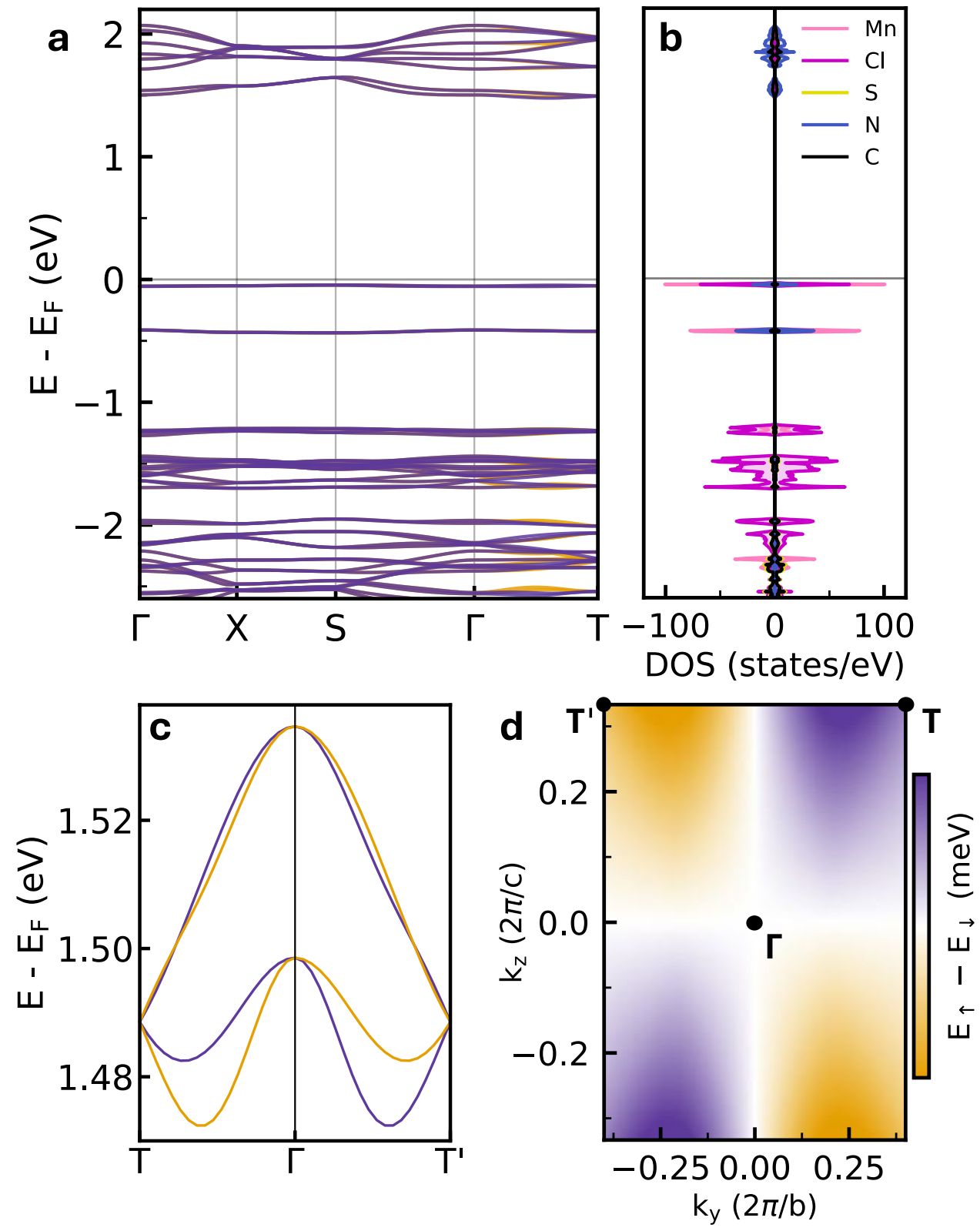

**Figure 3. a** Electronic band structure and **b** PDOS in $\mathbf{MnCl_2(tdz)_2}$. **c** depicts spin splitting along the *b//c* direction and **d** shows the full spin splitting along the *bc* plane.

The strong directional dependence of the AM spin splitting naturally gives rise to an anisotropic spin-transport response.[29,30] In particular, the symmetry relating inequivalent spin sublattices confines the AM signature to the *bc* plane, enabling a spin-splitter response in which spin-conserving transport deflects spin-up and spin-down carriers into opposite transverse channels. To verify this prediction, we perform Boltzmann transport calculations, revealing spin-dependent conductivity exclusively within the *bc* plane while remaining absent along the other crystallographic directions (Figure S12 and S13). This generates a spin accumulation on the opposite sides of the *bc* plane (Figure S4). The resulting spin-splitter response therefore constitutes the transport fingerprint of the anisotropic AM electronic structure in $\mathbf{MnX_2(tdz)_2}$.

The anisotropic electronic and transport properties in $\mathbf{MnX_2(tdz)_2}$ ultimately rely on the stability of the $AFM_3$ ground state. To understand its microscopic origin, we evaluated the magnetic exchange interactions ($J$) governing the spin Hamiltonian (Supplementary Note 2). Three exchange pathways were considered, corresponding to the first-neighbor intralayer interaction ($J_1$) and the first- and second-neighbor interlayer couplings ($J_2$ and $J_3$), respectively.

As summarized in Table 2, all exchange interactions are AFM, consistent with the dominant superexchange expected between HS $Mn^{2+}$ centers. The $AFM_3$ ground state is primarily stabilized by the cooperative action of $J_1$ and $J_2$, whereas the weaker AFM $J_3$ couples second-neighbor spins that are ferromagnetically aligned, introducing a minor magnetic frustration that does not overcome the dominant exchange network.

Both compounds exhibit in-plane magnetic anisotropy, with energies comparable to those of other layered van der Waals magnets.[31] As expected, magnetic anisotropy increases in **$MnBr_2(tdz)_2$** owing to the stronger SOC associated with Br atoms. Atomistic spin dynamics simulations predict Néel temperatures ($T_N$) of approximately 30 K (25 K) for **$MnCl_2(tdz)_2$** (**$MnBr_2(tdz)_2$**) (Figure S14), in good agreement with experimentally reported values for related layered magnetic MOFs.[32,33] Together, these results demonstrate that the $AFM_3$ ground state is magnetically robust, providing the stability required for the AM electronic structure and spin-splitter response described above.

**Table 2.** Calculated magnetic exchange interactions ($J$, in meV) and magnetic anisotropy energy ($D$, in meV/Mn atom).

| | $J_1$ | $J_2$ | $J_3$ | $D$ |
|---|---|---|---|---|
| **$MnCl_2(tdz)_2$** | -0.89 | -0.77 | -0.39 | 0.31 |
| **$MnBr_2(tdz)_2$** | -0.69 | -0.61 | -0.31 | 0.70 |

Having established the microscopic origin of the AM ground state, we now investigate whether it can be dynamically controlled by external stimuli. We apply hydrostatic pressure up to 10 GPa to **$MnX_2(tdz)_2$**. Our results reveal a sharp HS-to-low-spin (LS) transition above ~5 GPa. As shown in Figure 4a, the Mn magnetic moment abruptly decreases from ~4.5 $\mu_B$ to ~1 $\mu_B$, providing clear evidence of SCO. This transition is accompanied by a pronounced collapse of the unit-cell volume (Figure 4b), consistent with the reduced ionic radius of LS $Mn^{2+}$. Similar pressure-induced SCO transitions have been reported in octahedrally coordinated Mn compounds, although typically at considerably higher pressures (≈20–30 GPa).[34–37] The substantially lower transition pressure predicted here is attributed to the structural flexibility of the organic framework, which facilitates the electronic reconfiguration.[38,39]

The pressure-induced SCO transition reconfigures the Mn $d$-orbital occupation, modifying magnetic exchange interactions (Figure 4c). More broadly, pressure can alter molecular magnetic functionality through changes in the local coordination environment, crystal-field anisotropy and molecular vibrations, as demonstrated in single-molecule magnets.[40,41] Before the transition, hydrostatic pressure strengthens the dominant AFM exchange interactions ($J_1$ and $J_2$), reinforcing the stability of the $AFM_3$ ground state and raising $T_N$ to approximately 55 K (Figure S15). Thus, the nonrelativistic spin splitting is enhanced by nearly 100%, together with the associated spin-splitter response (Figure S16-S18). Therefore, moderate pressure does not merely preserve the AM phase but actively reinforces its electronic and transport functionalities.

Beyond the SCO, the pressure response changes qualitatively. The depopulation of the Mn $e_g$ orbitals suppresses the dominant superexchange pathways present in the HS configuration, leaving only weaker $t_{2g}$-mediated interactions.[42] Consequently, all $J$ decrease abruptly. More importantly, the SCO stabilizes a different magnetic ground state: $AFM_1$ becomes energetically preferred over $AFM_3$ (Figure 4d). Unlike $AFM_3$, the $AFM_1$ configuration preserves the $[C_2 \parallel i]$ symmetry operation relating opposite-spin sublattices, thereby restoring the symmetry that forbids altermagnetism. Accordingly, the electronic band structure becomes fully spin degenerate above the transition pressure, and the spin-splitter response disappears (Figure S19).

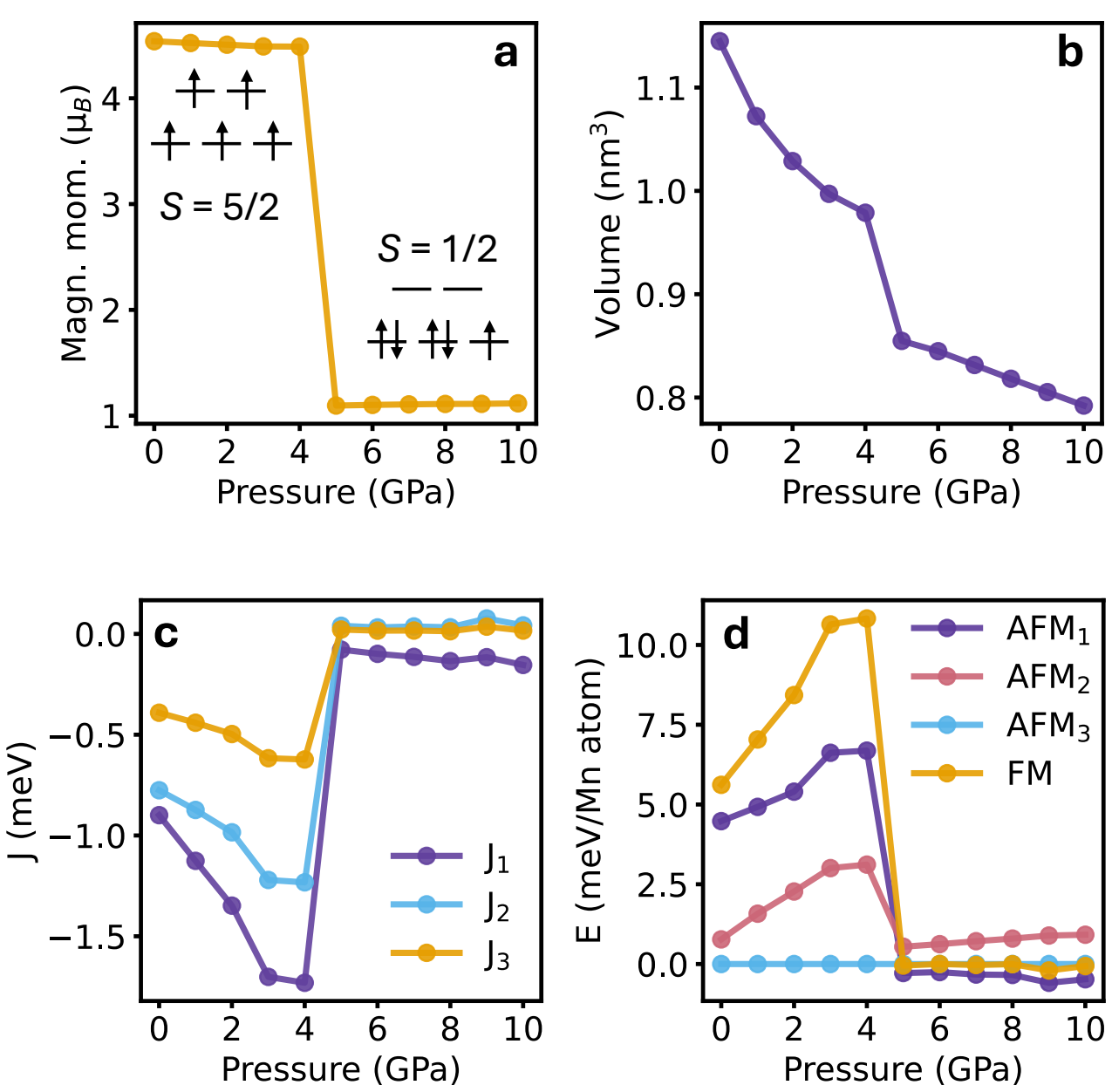

**Figure 4.** Effect of hydrostatic pressure on **a** magnetic moment in Mn atoms, **b** unit-cell volume, **c** magnetic exchange couplings and **d** energy of each magnetic configuration in **$MnCl_2(tdz)_2$**.

These results demonstrate that the cancellation of spin splitting is not a direct consequence of hydrostatic pressure itself, but rather of the pressure-induced SCO transition that changes the symmetry of the magnetic ground state. Because SCO can also be triggered by temperature, light, or other external stimuli in molecular materials, with oriented external electric fields offering a route to directly tune the spin-transition energetics,[43] this mechanism extends well beyond pressure control, providing a versatile approach to design switchable altermagnetic materials. Rather than simply modifying magnetic interactions, SCO changes the symmetry of the magnetic ground state, thereby switching the AM functionality itself.

In summary, we demonstrate that SCO provides an effective mechanism to dynamically switch altermagnetism in molecular materials. Using the layered van der Waals MOFs $\mathbf{MnX_2(tdz)_2}$ as a proof of concept, we show that anisotropic interlayer exchange stabilizes an AM ground state characterized by *d*-wave momentum-space spin splitting and a corresponding spin-splitter transport response. Under hydrostatic pressure, a HS to LS transition reconfigures the Mn *d*-orbital occupation, modifies the magnetic exchange network and stabilizes a different AFM ground state whose symmetry suppresses the nonrelativistic spin splitting.

These results demonstrate that the disappearance of altermagnetism is not a direct consequence of pressure itself, but rather of the SCO-driven change in magnetic symmetry. The reversibility of pressure-induced SCO has been demonstrated in 2D coordination polymers, where pressure cycling recovers the initial spin state and gives rise to piezohysteresis.[44,45] More broadly, this work establishes molecular altermagnets exhibiting SCO as a versatile platform to achieve dynamically switchable altermagnetism. Our results introduce a design principle for externally controllable spin-dependent functionalities and pave the way for adaptive, programmable molecular spintronic devices.

*Computational Details*. Structural relaxations, electronic structure calculations, magnetic configurations and ab initio molecular dynamics (AIMD) simulations were performed using the Vienna Ab initio Simulation Package (VASP).[46] The exchange-correlation energy was described within the generalized gradient approximation (GGA) using the Perdew–Burke–Ernzerhof (PBE) functional. Grimme's D3 dispersion correction was included to account for the weak van der Waals interactions between adjacent MOF layers. The partially occupied Mn 3d orbitals were treated within the GGA+U approach using an effective Hubbard parameter $U_{eff} = U - J = 3$ eV. Hybrid HSE06 calculations were additionally performed as a computational validation of the relative stability of the magnetic configurations and the predicted AM spin splitting. The projector augmented-wave (PAW) method was employed with a plane-wave energy cutoff of 500 eV. Brillouin-zone integrations were carried out using a Monkhorst–Pack k-point mesh of 5 × 5 × 3. The electronic self-consistency criterion was set to $10^{-7}$ eV, and all structures were fully relaxed until the residual forces on each atom were below 0.01 eV $Å^{-1}$. AIMD simulations were carried out within the canonical (NVT) ensemble for 5 ps using a time-step of 1 fs at 300 and 600 K. Hydrostatic pressure was simulated as an energy correction to the stress tensor. Néel temperatures were evaluated via atomistic spin simulations as implemented in Vampire software.[47] A sample of 50 x 50 x 50 nm was used in combination with 10000 equilibration and loop time steps. Maximally localized Wannier functions were constructed using Wannier90 code,[48] considering the d orbitals of Mn, the p orbitals of C, N, Cl and S, and the s orbitals of H atoms as the basis set. Transport calculations were performed within Boltzmann formalism, as implemented in BoltzWann code.[49] The scattering rate was set to Gamma = 1 meV and the temperature to 20 K.

## ASSOCIATED CONTENT

**Supporting Information**. Symmetry analysis; Spin Hamiltonian: Spin accumulation due to spin-splitter effect; Pressure-induced SCO; Additional information.

## AUTHOR INFORMATION

### Corresponding Author

* j.jaime.baldovi@uv.es

### Author Contributions

The manuscript was written through contributions of all authors. This work is part of the PhD thesis of D.L-A. All authors have given approval to the final version of the manuscript.

## ACKNOWLEDGMENTS

We thank Matthew J. Cliffe for fruitful discussions. The authors acknowledge financial support from the European Union (ERC-2021-StG 101042680 2D-SMARTiES), the Spanish MCIU (PID2024-162182NA-I00 2D-MAGIC), Spanish MICINN (Excellence Unit “Maria de Maeztu” CEX2024-001467-M) and the Generalitat Valenciana (grant CIDEXG/2023/1).